\documentclass{aa}  

\usepackage{graphicx}
\usepackage{verbatim}
\usepackage{txfonts}
\usepackage[dvipsnames]{xcolor}
\usepackage{hyperref}
\hypersetup{
    colorlinks=true,
    linkcolor=blue,
    filecolor=magenta,
    urlcolor=blue,
    citecolor=blue
}
\usepackage{comment}

\newcommand{\teff}{T_{\mathrm{eff}}} 
\newcommand{\lgg}{\log{g}} 
\newcommand{\vsini}{\varv\,\sin i} 
\newcommand{\vmic}{V_{\mathrm{turb}}} 
\newcommand{\lggf}{\log{gf}} 
\newcommand{\feh}{\mathrm{\left[Fe/H\right]}} 
\newcommand{\tife}{\mathrm{\left[Ti/Fe\right]}} 
\newcommand{\cafe}{\mathrm{\left[Ca/Fe\right]}} 
\newcommand{\tih}{\mathrm{\left[Ti/H\right]}} 
\newcommand{\cah}{\mathrm{\left[Ca/H\right]}} 
\newcommand{\kms}{\mathrm{km\,s^{-1}}} 

\begin{document} 

   \title{Abundance analysis of benchmark M dwarfs}

   \author{T. Olander\inst{\ref{inst:UAO}}
          \and
          U. Heiter\inst{\ref{inst:UAO}}
          \and
          N. Piskunov\inst{\ref{inst:UAO}}
          \and
          J. K{\"o}hler\inst{\ref{inst:TLS}}
          \and
          O. Kochukhov\inst{\ref{inst:UAO}}
          }

   \institute{
   Observational Astrophysics, Department of Physics and Astronomy, Uppsala University, Box 516, SE-751 20 Uppsala, Sweden, \email{ulrike.heiter@physics.uu.se}
   \label{inst:UAO}
   \and
   Thüringer Landessternwarte, Sternwarte 5, 07778 Tautenburg, Germany
   \label{inst:TLS}
   }

   \date{Received 17 December 2024; accepted 5 May 2025}
 
  \abstract
   {Abundances of M dwarfs, being the most numerous stellar type in the Galaxy, can enhance our understanding of planet formation processes. They can also be used to study the chemical evolution of the Galaxy, where in particular $\alpha$-capture elements play an important role.}
   {We aim to obtain abundances for Fe, Ti, and Ca for a small sample of well-known M dwarfs for which interferometric measurements are available. These stars and their abundances are intended to serve as a benchmark for future large-scale spectroscopic studies.}
   {We analysed spectra obtained with the GIANO-B spectrograph. Turbospectrum and the wrapper TSFitPy were used with MARCS atmospheric models in order to fit synthetic spectra to the observed spectra. We performed a differential abundance analysis in which we also analysed a solar spectrum with the same method and then subtracted the derived abundances line-by-line. The median was taken as the final abundance for each element and each star.}
   {Our abundances of Fe, Ti, and Ca agree mostly within uncertainties when comparing with other values from the literature. However, there are few studies to compare with.}
   {}


   \maketitle

\section{Introduction}
Recent years have shown an increased interest in M~dwarfs as they have become favourable targets in the hunt for exoplanets. This is because exoplanets are easier to detect around M~dwarfs due to the low mass and faintness of these cool stars, which favours both the radial velocity and the transit methods. Additionally, the proximity between star and potential planet in the habitable zone increases opportunities for characterising the planet by using, for example, transit spectroscopy \citep[cf. e.g.][]{2021A&A...645A..41L,2022AJ....164...96C,2024ESS.....562522L}. 

These cool and faint stars are the most common stars in the galaxy as over 70\% of the stars in our galaxy are M dwarfs \citep{Henry2006NrMdwarfs}. The multitude of M~dwarfs could make them important in characterising large stellar structures such as clusters. We are however still limited by our observational capabilities. M~dwarfs could also be important for tracing chemical evolution because of their long expected lifetime and slow evolution.
They would be an indicator of the local chemical makeup from when a group was formed. The difficulty would be to estimate the M~dwarf age but this problem
could be solved by studying stars in clusters. The multitude of stars also increases the likelihood of finding a habitable exoplanet as it is estimated that each M~dwarf has over two Earth-like exoplanets \citep{Dressing2015_MdwarfsPlanets}.

M dwarfs are however difficult targets to observe. The cool temperatures in the photosphere of M dwarfs makes it possible for di-atomic and tri-atomic molecules to form, such as TiO, FeH, H$_2$O, VO and more  \citep{Grey2009StellarSpectralClass}. Atomic lines are hidden and blended by these molecular lines. This is especially true in the optical wavelength region where bands of TiO blend with the atomic lines. In the near-infrared (NIR) we find less strong molecular features and the atomic lines are easier to distinguish, but not without issue. For example, in the H-band we find a lot of lines of water molecules which suppress the continuum, forming a pseudo-continuum \citep{APOGEE_Sarmento2021}. There is less water in the J-band but instead telluric lines are imprinted on the spectra when using ground-based observations. One can either ignore the telluric lines and use atomic lines in-between, observe a telluric standard star or use software (as explained later) in order to remove the telluric lines. 

Despite these challenges multiple spectroscopic studies have determined effective temperature, metallicity, and surface gravity. A few examples of such studies are \citet{2015ApJMann},  \citet{Lind2016,Lind2017}, \citet{Pass2018,Pass2019}, \citet{Rajpurohit2018CARMENES}, \citet{APOGEE_Sarmento2021}, and \citet{2022ApJSouto}. Several of these studies used either calibrations or photometry in order to obtain the surface gravity because of the degeneracy found between the parameters. 

It is important to obtain abundances of stars in general, in order to explore and understand galactic chemical evolution. In particular, abundances of $\alpha$-capture elements allow one to determine the relative contributions of the end stages of different types of masses and thereby the nucleosynthetic history of stellar populations within the Milky Way.
Abundances of M~dwarfs will also lead to progress in our understanding of planet formation due to the increased number of exoplanets detected around M~dwarfs.
Key elements in this regard are C, S, O, Na, Si, Fe, Mg, Ni, Ca, and Al (in order from most volatile to most refractory, \citealt{2019MNRAS.482.2222W}). 
By determining the abundances of these elements in a host star and accounting for the depletion of volatiles during the formation of rocky planets the interior compositions and structures of terrestrial exoplanets can be constrained and compared to the Earth.

The ratios of the elemental abundances found in the star are often assumed to be mirrored in the planet \citep[e.g. ][]{2015A&A...580A..30T,2024AJ....168..281B}. 
However, for some high-density planets indications for differences between the abundances of the planet and the host star have been found \citep[e.g. ][]{2018NatAs...2..393S}. 
These cases may be used to explore the special conditions required to form Mercury-like planets.
Few abundance studies of M~dwarfs have been published up to now. Here is a sample with the derived species in parentheses: \citet{2005MNRAS.356..963Woolf} (Fe and Ti), \citet{2016PASJ...68...13Tsuji} (C and O), \citet{2020AbiaSrRb} (Rb, Sr, and Zr ), \citet{2020AAMaldonado} (C, O, Na, Mg, Al, Si, Ca, Sc II, Ti, V, Cr, Mn, Co, Ni, Zn), \citet{2021YutongShan} (V), \citet{2022ApJSouto} (C, O, Na, Mg, Al, Si, K, Ca, Cr, Mn, Fe, and Ni),  \citet{2022AJ_Ishikawa} (Na, Mg, Si, K, Ca, Ti, V, Cr, Mn, Fe, and Sr), \citet{2024ApJ_Melo} (Fe, C, O, Na, Mg, Al, Si, K, Ca, Ti, V, Mn), \citet{2024arXiv_Jahandar} (Fe, Mg, O, Si, Ca, Ti, Al, Na, C, and K).

When performing abundance analyses it is important to have an accurate effective temperature and surface gravity. In this work we used data obtained with interferometry. With an angular diameter from interferometry and a bolometric flux from photometry, the Stefan-Boltzmann law can be used in order to obtain model-independent effective temperatures. Using distances from astrometry and empirical  mass-luminosity relations, the mass and thereby the surface gravity of the star can be obtained. Some interferometric studies of M~dwarfs which have been published up to date include \citet{2006ApJ...644..475B, 2011ApJ...729L..26V, Boyajian2012, 2014MNRAS.438.2413V, 2016A&A...593A.127K, 2018ApJ...858...71S, Rabus2019, 2022A&A...665A.120C, Ellis2022}. 

Many current and future surveys include M~dwarfs in their sample, which means that thousands of M~dwarfs need to be accurately characterised. APOGEE \citep{2017AJ_APOGEE,2020AJ_Apogee} is a survey primarily targeting red giants but they have also observed M~dwarfs. Another example is PLATO \citep{2014ExARauer_PLATO,2024arXiv240605447R_plato}, which is a future space telescope that will use light curves in order to detect exoplanets. Thousands of M~dwarfs are part of the PLATO sample, and spectra from follow-up observations need to be analysed.
New methods using machine learning have been developed in order to facilitate the analysis of large samples of stars such as these. Some examples of such studies are \citet{2020ApJ_Birky,2020A&A_Passegger,2022AaA_Passegger,2024MNRA_Rains}, and \citet{Olander2025}.
In some cases the observed spectra of well known stars are used directly in the training and other times they are used in order to verify the results.
Due to the difficulties discussed above there has been a lack of well characterised M~dwarfs. We have effective temperatures and surface gravity obtained from interferometry for a set of M~dwarfs but we also need the metallicity and abundances of various elements, especially the $\alpha$-elements. This is obtained by carefully analysing high-resolution spectra of individual stars, which has been done by the studies referenced above.  

In this article we present an abundance analysis of a sample of M~dwarfs with interferometric measurements from \citet{Boyajian2012}. In Sect.~\ref{sec:sample-obs} we introduce our sample and observations. In Sect.~\ref{sec:Method} we describe our analysis method. In Sect.~\ref{sec:result} we present our results followed by a discussion and conclusions in Sect.~\ref{sec:discussion}.

\section{Sample and observations}
\label{sec:sample-obs}
The sample of stars presented in this paper consists of well characterised stars with $\teff$ and $\lgg$ obtained using interferometry in \citet{Boyajian2012}. They obtain $\teff$ by measuring the limb-darkened angular diameter with the CHARA array and using the bolometric flux obtained from Hipparcos parallaxes \citep{2007A&A...474..653V} and photometry fitted to spectral templates in the Stephan-Boltzmann law. The surface gravity is obtained using an absolute K-band mass-luminosity relation from \citet{1993AJ....106..773H}. 
The sample consists of nine stars and includes one binary system. The stars are from spectral type M1 to M4 and in the literature the majority have a sub-solar metallicity.
The sample can be seen in Table \ref{tab:sample} together with the total exposure time and the signal-to-noise ratio (SNR).
The SNR is given at 11785~\AA, which corresponds to the centre of the part of the spectra used in this article (centre of order 65, see Sect.~\ref{sec:obs:reduc}).
In case of multiple observations per star (two for GJ~412A, GJ~699, and GJ~725A; four for GJ~436), which were co-added after data reduction during telluric removal (see Sect.~\ref{sec:obs:tell}), the SNR is the square root of the sum of the squared SNRs of each observation.

\begin{table*}[t]
  \caption{Target sample.}
  \label{tab:sample}
  \centering
  \begin{tabular}{lllllcrc}
  \noalign{\smallskip}
  \hline
  \noalign{\smallskip}
  \text{Name} & \text{Simbad identifier} & \text{Sp type} & \text{RA} & \text{DEC} & \text{K mag} & \text{Exp. time [s]} & \text{SNR} \\
  \noalign{\smallskip}
  \hline
  \noalign{\smallskip}
  GJ 411 & HD 95735 & M2+V & 11 03 20.1948 & +35 58 11.576 & 3.340 & 120 & 281 \\
  GJ 412A & BD+44 2051 & M1.0V & 11 05 28.5769 & +43 31 36.386 & 4.769 & 600 & 293 \\
  GJ 436 & Ross 905 & M3V & 11 42 11.0933 & +26 42 23.650 & 6.073 & 1600 & 274 \\
  GJ 526 & HD 119850 & M2V & 13 45 43.7755 & +14 53 29.471 & 4.415 & 200 & 216 \\
  GJ 581 & BD$-$07 4003 & M3V & 15 19 26.8269 & $-$07 43 20.189 & 5.837 & 600 & 181 \\
  GJ 699 & Barnard's star & M4V & 17 57 48.4984 & +04 41 36.113 & 4.524 & 320 & 190 \\
  GJ 725A & HD 173739 & M3V & 18 42 46.7043 & +59 37 49.409 & 4.432 & 240 & 172 \\
  GJ 725B & HD 173740 & M3.5V & 18 42 46.8946 & +59 37 36.721 & 5.000 & 600 & 224 \\
  GJ 809 & HD 199305 & M1.0V & 20 53 19.7889 & +62 09 15.817 & 4.618 & 400 & 210 \\
  \noalign{\smallskip}
  \hline
  \end{tabular}
  \tablefoot{Data were obtained from Simbad using the Python module astroquery.simbad. Exp. time is total time on target (both A and B nodding positions and sum of individual exposures in the case of multiple observations). See text for specification of signal-to-noise ratio (SNR).}
\end{table*}

\subsection{Observational data}
\label{sec:obs}  
The observations were performed with the GIANO-B spectrograph mounted on the 3.58~m Telescopio Nazionale Galileo on La Palma. GIANO-B is a near-infrared echelle spectrograph with a resolving power of 50~000 and a wavelength coverage from 9000 to 24~500~\AA\ \citep{Origlia2014_GIANO-B}. Observations were conducted on 25 to 27 May 2019 in visitor mode. The observations were planned with the goal of achieving an SNR of at least 150.

\begin{figure}
    \centering
    \includegraphics[width=1.0\linewidth]{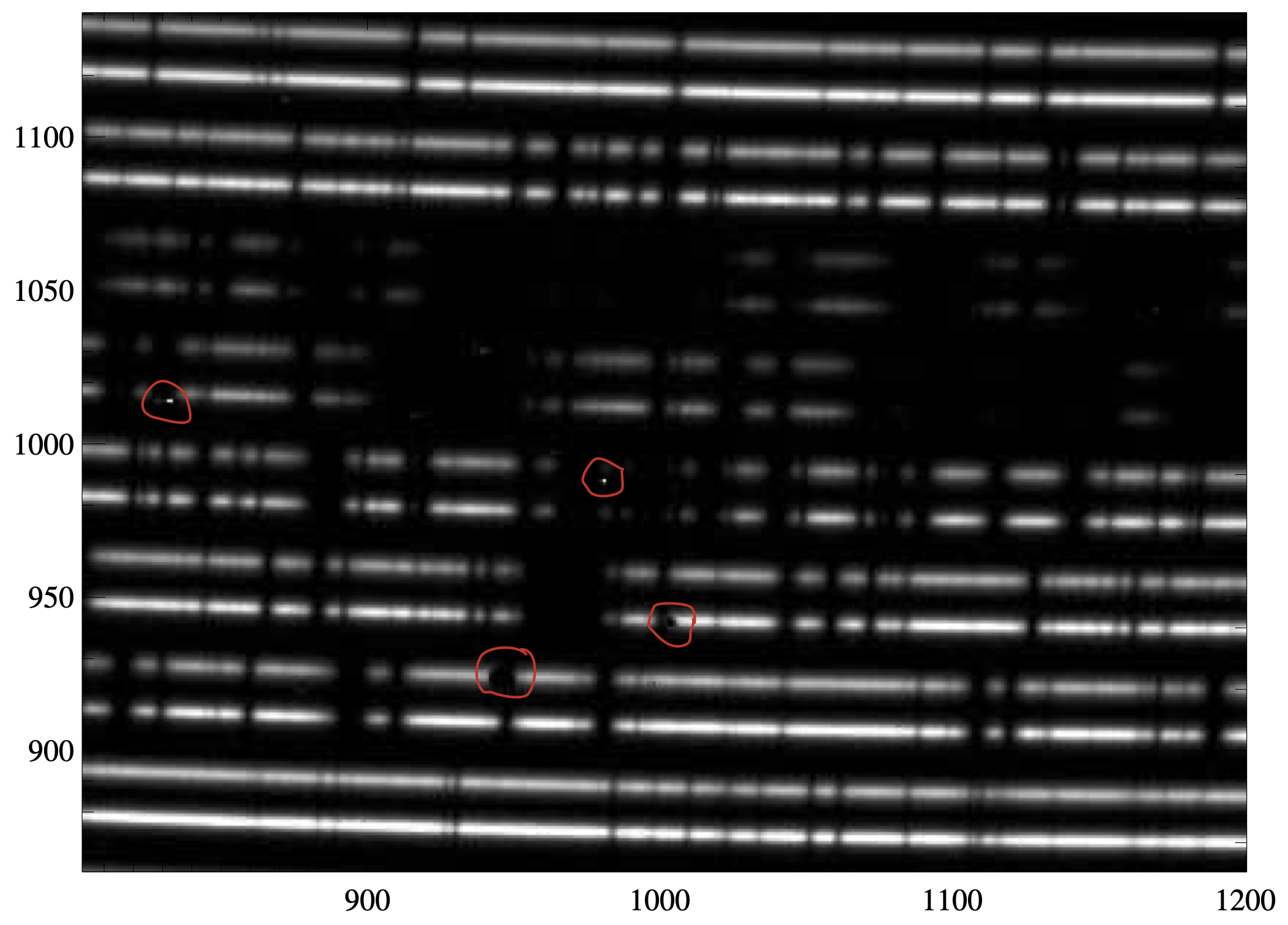}
    \caption{A fragment of spectra taken at two nodding positions by GIANO-B after combining them.
    One can see replicated spectral lines corresponding to the same spectral orders, the tilt of the slit image, a couple of strong cosmic ray hits and some detector defects outlined in red.}
    \label{fig:GIANO raw fragment}
\end{figure}

\subsection{Data reduction}
\label{sec:obs:reduc}
The focal plane format of GIANO-B represents a challenge for data reduction due to 50 curved spectral orders present in two nodding positions, multiple detector defects, variable tilt of the slit images, etc., as illustrated in Fig.~\ref{fig:GIANO raw fragment}, where artefacts are marked in red. We have selected to use the REDUCE package \citep{2002A&A...385.1095P,2021A&A...646A..32P}, 
which is well tested on several instruments. This includes the ESO VLT CRIRES+ spectrograph, which has a number of similarities with GIANO-B. After constructing a master flat, tracing spectral orders, and extracting the wavelength calibrations (UNe calibration lamp) we combined the raw frames from the A and B nodding positions and extracted them as one frame, accounting for the slit tilt. The slit orientation was mapped from the UNe frames and then used for the optimal extraction of the science spectra as described in \citet{2021A&A...646A..32P}. Continuum normalisation in REDUCE requires wavelength overlap between orders, which forced us to restrict the data reduction to the 36 bluest orders (9400 to 17000~\AA), and here we encountered the biggest challenge in our reduction. The usual initial guess for the continuum fit required for order splicing is the blaze functions, derived from the master flat. These show pronounced fringing with an amplitude of a few percent and a frequency comparable to the width of the spectral lines. To resolve this problem we first fitted a ``continuum'' to the blaze functions not affected by telluric absorption, then fitted a 2D surface to the upper envelope of the spectral format by associating each blaze point with its position on the detector and interpolating to the orders affected by tellurics. The corrected set of blaze functions was good enough to produce a decent continuum fit for orders 45 to 80 that are used in this article.

\subsection{Telluric removal}
\label{sec:obs:tell}
Telluric lines were removed using an updated version of the software {\tt viper}\footnote{\url{https://mzechmeister.github.io/viper_RV_pipeline}} (Velocity and IP EstimatoR, \citealt{2021ascl_viper}).
The main concept of {\tt viper} is based on the method described by \citet{1996PASP..108..500B}. The algorithm has been extended by adding a term for the telluric spectrum, which enables its usage for observations taken in the NIR. Using least-squares fitting, the wavelength solution, blaze, instrument profile and telluric line depths of the observations are modelled simultaneously to account for instrument instabilities and weather changes. Although {\tt viper} was originally developed for the determination of radial velocities (RVs), it can also be used for pure telluric correction. For the telluric forward modelling {\tt viper} uses synthetic standard models obtained from Molecfit \citep{2015A&A...576A..77S_Molecfit} for each molecule in the Earth atmosphere. For details regarding the telluric removal the reader is directed to \citet{2024_viper}.

We fed {\tt viper} with pre-normalised spectra and adjusted the settings in order to obtain the best possible telluric removal. Still we faced problems in the removal of deep tellurics, which cannot be accurately modelled. However, we avoided spectral lines blended with or those that are near strong telluric lines in our analysis. In cases where we have multiple observations for the same star the spectra were co-added within {\tt viper}, by using a weighted average of all the telluric-corrected spectra.

\section{Method}
\label{sec:Method}
For the analysis we used Turbospectrum \citep{Turbospectrum_Alvarez1998,Turbo_2012ascl.soft05004P,Gerber_Turbo_2023A&A} together with the wrapper TSFitPy \citep{Gerber_Turbo_2023A&A,Storm_tsfitpy_2023MNRAS.525.3718S} and MARCS atmospheric models \citep{Gustafsson2008A&A}. It uses the \citet{2021MNRAS_Bergemann} and \citet{2022A&A_Magg} solar abundances. 
TSFitPy fits synthetic spectra generated by Turbospectrum to observed spectra using Nelder-Mead minimisation. In this work we use line-by-line fitting and thereby obtain abundances for each line. In order to minimise the effects from used models we perform a differential abundance analysis. We analyse the high resolution solar spectrum provided by \citet{Reiners2016_solar_atlas} using the same method as for the M~dwarfs. The one difference was that the solar metallicity was always set to zero and we did not alter the Ti and Ca abundances. The other parameters were set to $\teff$:~5771~K and $\lgg$:~4.44~dex\footnote{The units of surface gravity are $\mathrm{cm\,s^{-2}}$. However, throughout the article, we use the unit dex when specifying values of $\lgg$.}. We then subtract the solar abundance for each line from the M~dwarf abundance for the same line. The solar spectrum was obtained using a Fourier transform spectrograph (FTS) and the resolution in the NIR was $10^6$.

For the M~dwarf input parameters we used \citet{Boyajian2012} for $\teff$ as well as radius and mass to calculate $\lgg$. The exception is the binary GJ~725A and B where we used \citet{2015ApJMann}. There is an inconsistency for the parameters from \citet{Boyajian2012} for the B star. The effective temperature for that star differs significantly from other measurements which indicates a possible issue with the measured angular diameter. This has been seen for example in \citet{2020ApJ_Souto} and \citet{Olander2025}. For projected equatorial rotational velocity $\vsini$ and radial velocity RV we used \citet{2018A&AReiners} and \citet{2018MNRASFouque}. The rotational velocity was fixed throughout the analysis, and set to the upper limit given in the references. Our sample consists of slowly rotating stars and the effect from magnetic broadening should therefore be minimal. We note that our models do not include broadening caused by magnetic fields. The radial velocity is allowed to be adjusted slightly for a better fit. This is done by a second fitting after the abundance, see \citet{Storm_tsfitpy_2023MNRAS.525.3718S} for more details. For the results presented in this work the radial velocity rarely shifted by more than 2~km~s$^{-1}$.
For input metallicity $\feh$ we used \citet{2015ApJMann}. TSFitPy calculates the microturbulence $\vmic$ using an empirical relation based on $\teff$, $\lgg$, and $\feh$\footnote{$\vmic = 1.05 + 2.51\times10^{-4} (\teff - t0) + 1.5\times10^{-7} (5250 - t0)^2 - 0.14 (\lgg - g0) - 0.005 (\lgg - g0)^2 + 0.05 \feh + 0.01 \feh^2$, where $t0$ = 5500 and $g0$ = 4.}, which was derived for the data analysis within the Gaia-ESO Public Spectroscopic Survey (M. Bergemann, priv. comm.). We set the macroturbulence as free. The adopted input parameters are given in Table~\ref{tab:input_param}.

\begin{table*}[t]
   \caption{Stellar parameters used as input data.}
   \label{tab:input_param}
   \centering
   \begin{tabular}{lllrrrr}
   \noalign{\smallskip}
   \hline
   \noalign{\smallskip}
   \text{Star} & $\teff$ & $\lgg$ & $\feh$ & $\vsini$ & RV & Ref\\
               & [K] & [dex] & [dex] & [$\kms$] & [$\kms$] & \\
   \noalign{\smallskip}
   \hline
   \noalign{\smallskip}
   GJ 411 & 3465 & 4.856 & $-$0.38 & <2 & $-$84.835 & 1, 3  \\
   GJ 412A & 3497 & 4.843 & $-$0.37 & <2 & 68.760 & 1, 3  \\
   GJ 436 & 3416 & 4.796 & 0.01 & <2 & 9.482 & 1, 3  \\
   GJ 526 & 3618 & 4.784 & $-$0.31 & <2 & 15.685 & 1, 3  \\
   GJ 581 & 3442 & 4.959 & $-$0.15 & <2 & $-$9.530 & 1, 3   \\
   GJ 699 & 3224 & 5.059 & $-$0.40 & <2 & $-$110.579 & 1, 3  \\
   GJ 725A & 3441 & 4.871 & $-$0.23 & 2.1 & $-$0.580 & 2, 4  \\
   GJ 725B & 3345 & 4.960 & $-$0.30 & 2.4 & 1.190 & 2, 4  \\
   GJ 809 & 3692 & 4.719 & $-$0.06 & <2 & $-$17.320 & 1, 3  \\
   \noalign{\smallskip}
   \hline           
   \end{tabular}
   \tablebib{(1) \citet{Boyajian2012}; (2) \citet{2015ApJMann}; (3) \citet{2018A&AReiners}; (4) \citet{2018MNRASFouque}}
\end{table*}

\subsection{Line data and mask}
\label{sec:method:linedata}
The abundances determined in this work are based on carefully selected individual atomic lines most suitable for the purpose. However, for a realistic synthetic spectrum calculation including background lines and blends we need information on all transitions, as far as possible, contributing to the spectrum in the observed wavelength range in the stars of interest.
At the same time, for an efficient spectrum calculation we want to avoid treating unnecessary transitions.
Our line list was thus constructed to include atomic lines relevant for stellar parameters encompassing those of the target stars.
For this purpose, we used the VALD database \citep{VALD_1995A&AS..112..525P,VALD_2015PhyS...90e4005R,VALD_2019ARep...63.1010P} and its ``Extract Stellar'' mode.
The VALD extraction was done on 10 April 2024 using version 3735 of the VALD3 database and software and the default configuration.

We downloaded two line lists between 10~000 and 14~000~\AA, one for an effective temperature of 3000~K and a surface gravity of 4.5~dex, and another for $\teff$=4000~K and $\lgg$=5.5~dex.
We used the same minimum threshold for central line depth of 0.001\footnote{We note that the central line depth is estimated by the VALD software for extraction, without applying any instrumental or rotational broadening. This information is however not used in the further analysis.}, a microturbulence of 2~$\kms$ and elemental abundances enhanced by 0.5~dex for the extraction of both lists.
The lists were subsequently merged by removing duplicates.
Detailed information about those individual lines which were analysed including references can be found in Table~\ref{tab:linelist} in the appendix.
In addition to atomic data, we also included molecular data.
The data used for the molecular lines were the same as in \citet{Molecule_list_2012ApJ...751..156B}, and they were obtained from Bertrand Plez in private communication.
We note that for the calculation of the synthetic spectrum in the fitting process all of the atomic transitions extracted from VALD were used (about 1000 in the wavelength range covered by the analysed lines) together with the molecular data.

A line mask was created for each atomic species by visually inspecting an M~dwarf spectrum and the solar spectrum. The mask covers the deepest flux point of the observed line and stretches in general from the central wavelength given in the line list to the nearest local maximum to both left and right. Lines blended with telluric lines and other atomic lines were avoided. The cores of the included lines had to be clearly separated from other line cores, which was assessed on a line by line basis. The software iSpec \citep{2014A&A_iSPec,2019MNRAS_iSpec} was used for producing the line masks.
The lines used for each star and each element are listed in Tables~\ref{tab:fitted_Fe_lines} to Tables~\ref{tab:fitted_Ca_lines} in the appendix.

\subsection{Metallicity and abundances}\label{sec:method:metallicity}
We fit for the abundances of Fe, Ti, and Ca, one at a time in an iterative process as follows. We first fit for Fe in order to get the overall metallicity which is then fixed for the rest of the analysis. We then fit for Ti and thereafter Ca with the previously derived Ti fixed. We then go back to Ti with Ca fixed and the old Ti as an input parameter. We then analyse Ca again with the last Ti fixed and the previous Ca abundance as an input parameter. This iterative process was done in order to break possible degeneracies and dependencies between the abundances. Between each iteration we visually inspected the best fit synthetic spectra and discarded lines that showed a bad fit. This was done by first discarding observed lines containing artefacts. Then we removed lines where the derived abundance was outside of 1 to 2 standard deviations from the median value. Which tolerance was used depended on the number of available lines. If there were few lines we were more forgiving.
The final abundance in each iteration was obtained by taking the median of the abundances derived for the acceptable lines.
We obtained the uncertainty in the derived abundances by calculating the median absolute difference (MAD) between the different abundances for each individual line after subtracting the corresponding solar abundances.

\begin{figure*}
   \centering
   \includegraphics[width=17cm]{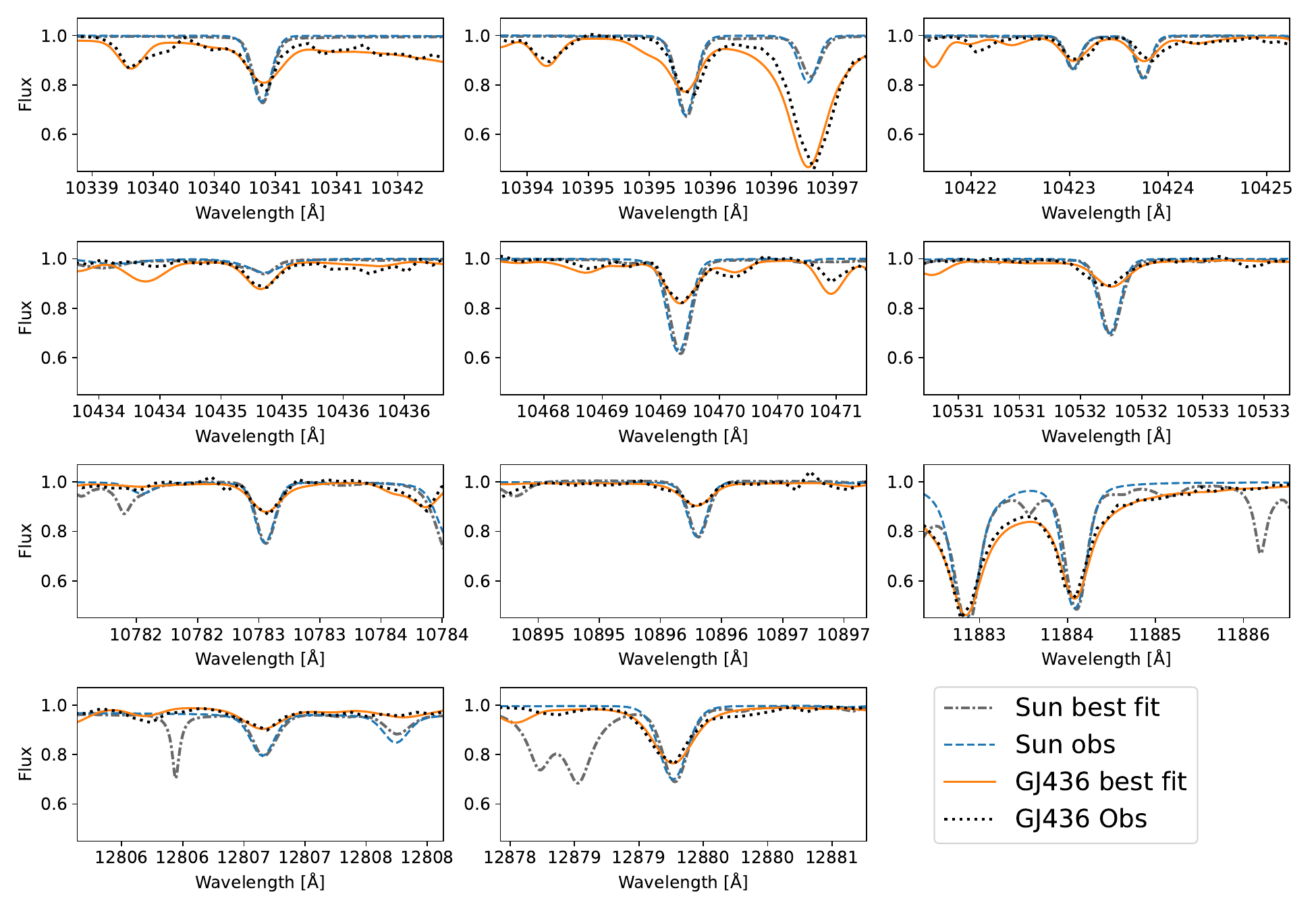}
   \caption{Best-fitted Fe~I lines in GJ~436 and the Sun. The observed spectra were only adjusted according to the literature radial velocity and not the RV fit for individual lines. The wavelength range 10\,421~\AA\ to 10\,425~\AA\ has two Fe lines. In the other subplots the Fe line is in the middle of the plot.}
   \label{fig:Fe_Gj436}
\end{figure*}

\begin{figure*}
   \centering
   \includegraphics[width=17cm]{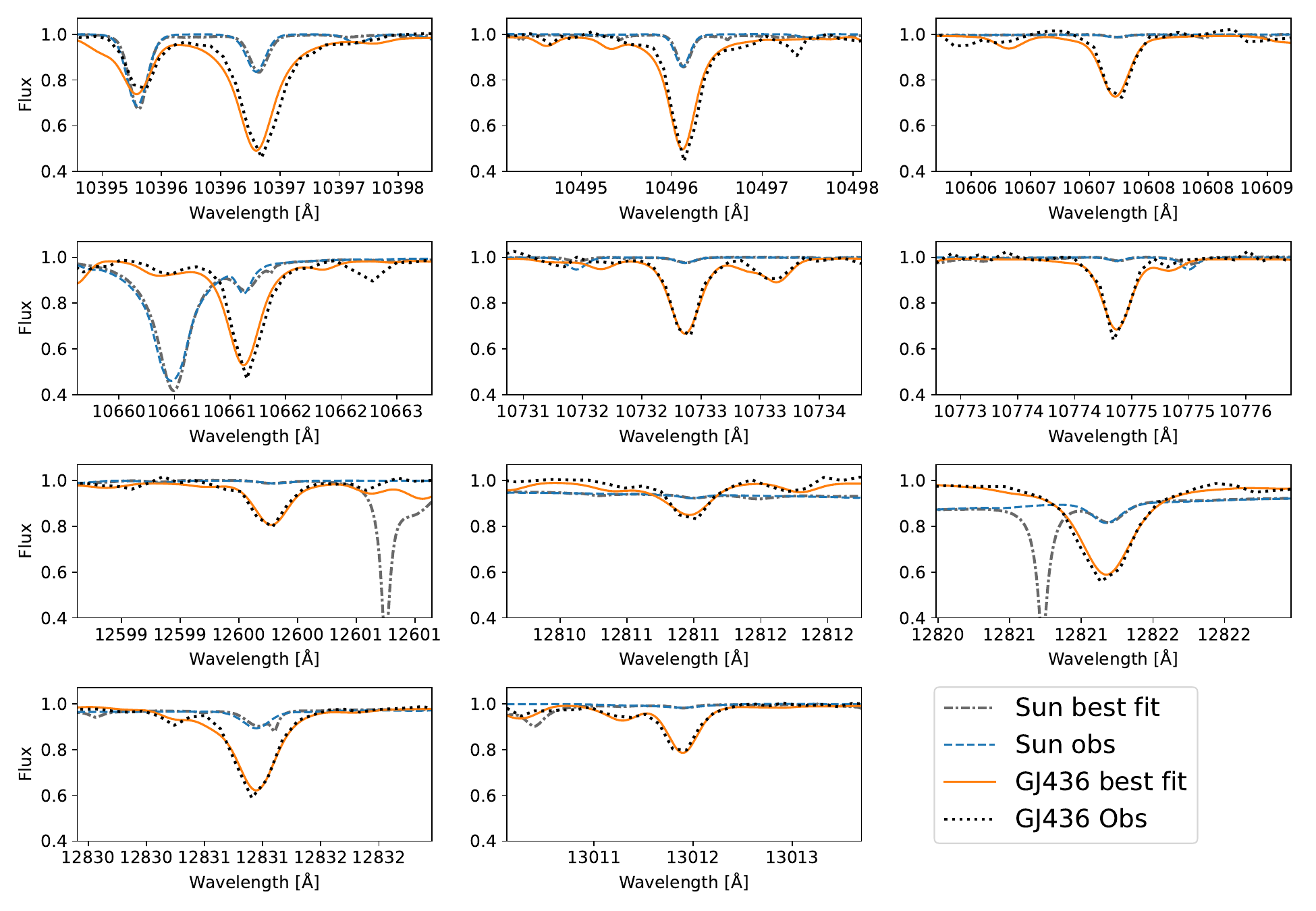}
   \caption{Same as Fig.~\ref{fig:Fe_Gj436}, for Ti~I lines.}
   \label{fig:Ti_Gj436}
\end{figure*}

\begin{figure*}
   \centering
   \includegraphics[width=17cm]{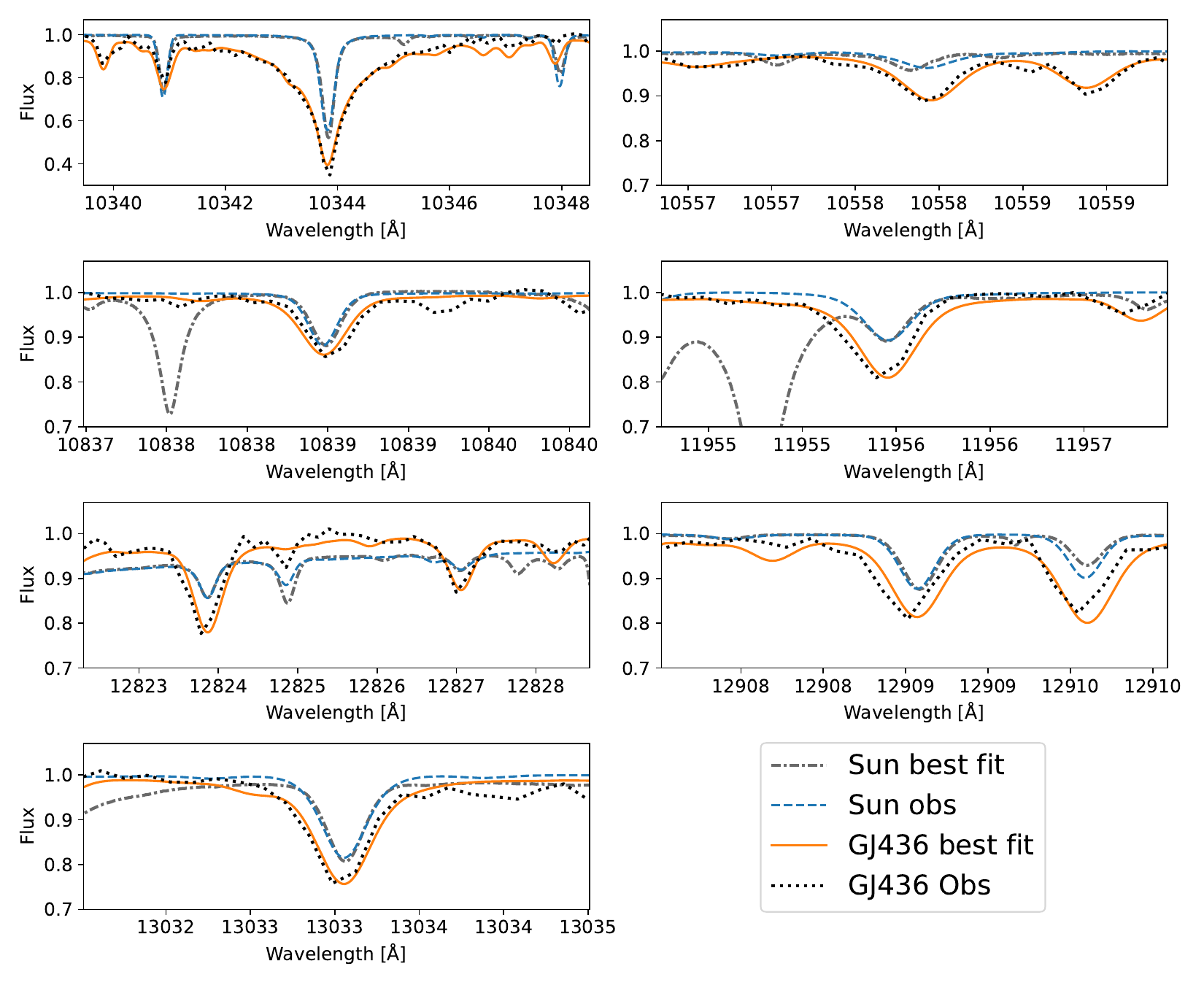}
   \caption{Same as Fig.~\ref{fig:Fe_Gj436}, for Ca~I lines. The wavelength range 12\,822~\AA\ to 12\,829~\AA\ has two Ca lines. In the other subplots the Ca line is in the middle of the plot.}
   \label{fig:Ca_Gj436}
\end{figure*}

\section{Results}
\label{sec:result}

\begin{table*}
\centering
\caption{Abundance ratios with uncertainties.}
\label{tab:result}
\begin{tabular}{lccccc}
\noalign{\smallskip}
\hline
\noalign{\smallskip}
Star & [Fe/H] & [Ti/Fe] & [Ca/Fe] & [Ti/H] & [Ca/H] \\
\noalign{\smallskip}
\hline
\noalign{\smallskip}
GJ 411  & $-0.373$ ± 0.143 & $+0.084$ ± 0.172 & $-0.079$ ± 0.193 & $-0.289$ ± 0.224 & $-0.452$ ± 0.240 \\
GJ 412A & $-0.451$ ± 0.209 & $+0.094$ ± 0.144 & $-0.027$ ± 0.180 & $-0.357$ ± 0.254 & $-0.478$ ± 0.276 \\
GJ 436  & $-0.010$ ± 0.131 & $+0.110$ ± 0.111 & $+0.052$ ± 0.108 & $+0.100$ ± 0.172 & $+0.042$ ± 0.170 \\
GJ 526  & $-0.162$ ± 0.183 & $+0.172$ ± 0.093 & $-0.094$ ± 0.153 & $+0.010$ ± 0.205 & $-0.256$ ± 0.239 \\
GJ 581  & $-0.151$ ± 0.259 & $+0.186$ ± 0.158 & $-0.012$ ± 0.084 & $+0.035$ ± 0.303 & $-0.163$ ± 0.272 \\
GJ 699  & $-0.582$ ± 0.182 & $-0.009$ ± 0.214 & $+0.086$ ± 0.133 & $-0.591$ ± 0.281 & $-0.496$ ± 0.225 \\
GJ 725A & $-0.477$ ± 0.259 & $+0.028$ ± 0.104 & $+0.041$ ± 0.062 & $-0.449$ ± 0.279 & $-0.436$ ± 0.266 \\
GJ 725B & $-0.329$ ± 0.202 & $+0.057$ ± 0.134 & $-0.067$ ± 0.075 & $-0.272$ ± 0.242 & $-0.396$ ± 0.215 \\
GJ 809  & $+0.221$ ± 0.158 & $+0.081$ ± 0.087 & $-0.195$ ± 0.054 & $+0.302$ ± 0.180 & $+0.026$ ± 0.167 \\
\noalign{\smallskip}
\hline
\end{tabular}
\tablefoot{Uncertainties (MAD) are given as $\pm$ values.}
\end{table*}

We present our results in Table~\ref{tab:result}. For some of the stars our uncertainties are high, this is due to a large spread in derived abundances from the different lines.
Which lines were used in the fitting procedure for each star can be seen in Appendix~\ref{append:fitted_lines}. An example of best fitted lines for Fe, Ti, and Ca can be seen in Figs.~\ref{fig:Fe_Gj436}, \ref{fig:Ti_Gj436}, and \ref{fig:Ca_Gj436}. In the figures we plot the output synthetic spectra from TSFitPy. We note that for the construction of the figures the observed spectra were only adjusted according to the literature radial velocity and not what the individual lines were fitted for, which is why they do not always line up. In some cases very weak lines in either the solar or M~dwarfs spectra were used in order to increase the number of used lines. The line had to have more than five data points and be clearly visible in comparison to the noise level. An example of this can be seen in Fig.~\ref{fig:Ti_Gj436} in the last panel were the line in the solar spectrum is really weak. The line is clear when zoomed in more. As can be seen in Figs.~\ref{fig:Fe_Gj436} to \ref{fig:Ca_Gj436} the fit of the lines in both the Sun and in the M~dwarf are good. In some cases the line core is not correctly fitted as can be seen for the Ti line at 10496~\AA\ or at 10661~\AA\ in Fig.~\ref{fig:Ti_Gj436} or the Ca line in Fig.~\ref{fig:Ca_Gj436} at 10344~\AA. This is an issue for some Ti and Ca lines. The cause could be incorrect atomic data or departure from local thermodynamic equilibrium, non-LTE, see more in Sect.~\ref{sec:discussion}.

\begin{figure*}
   \centering
   \includegraphics[width=5.8cm]{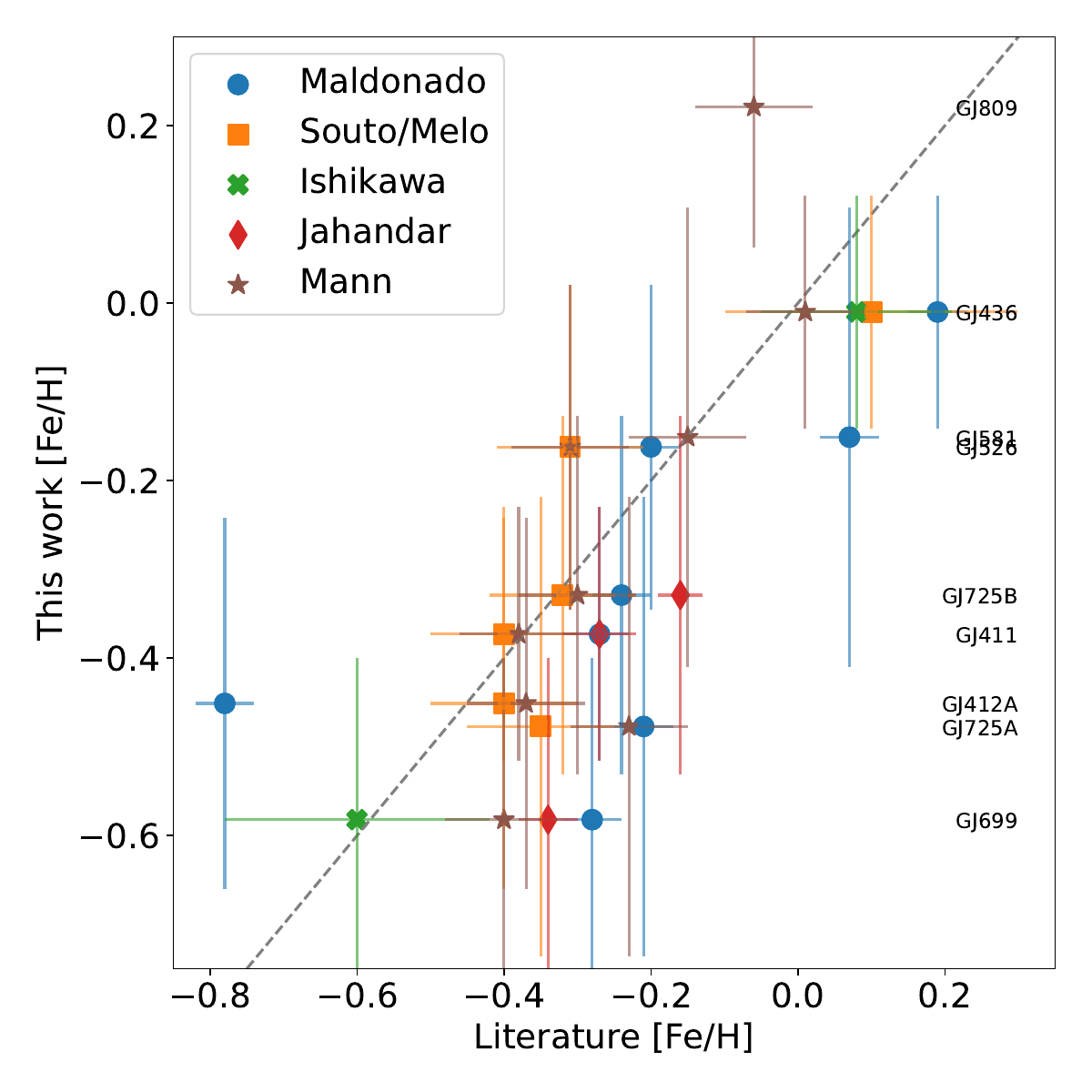}
   \includegraphics[width=5.8cm]{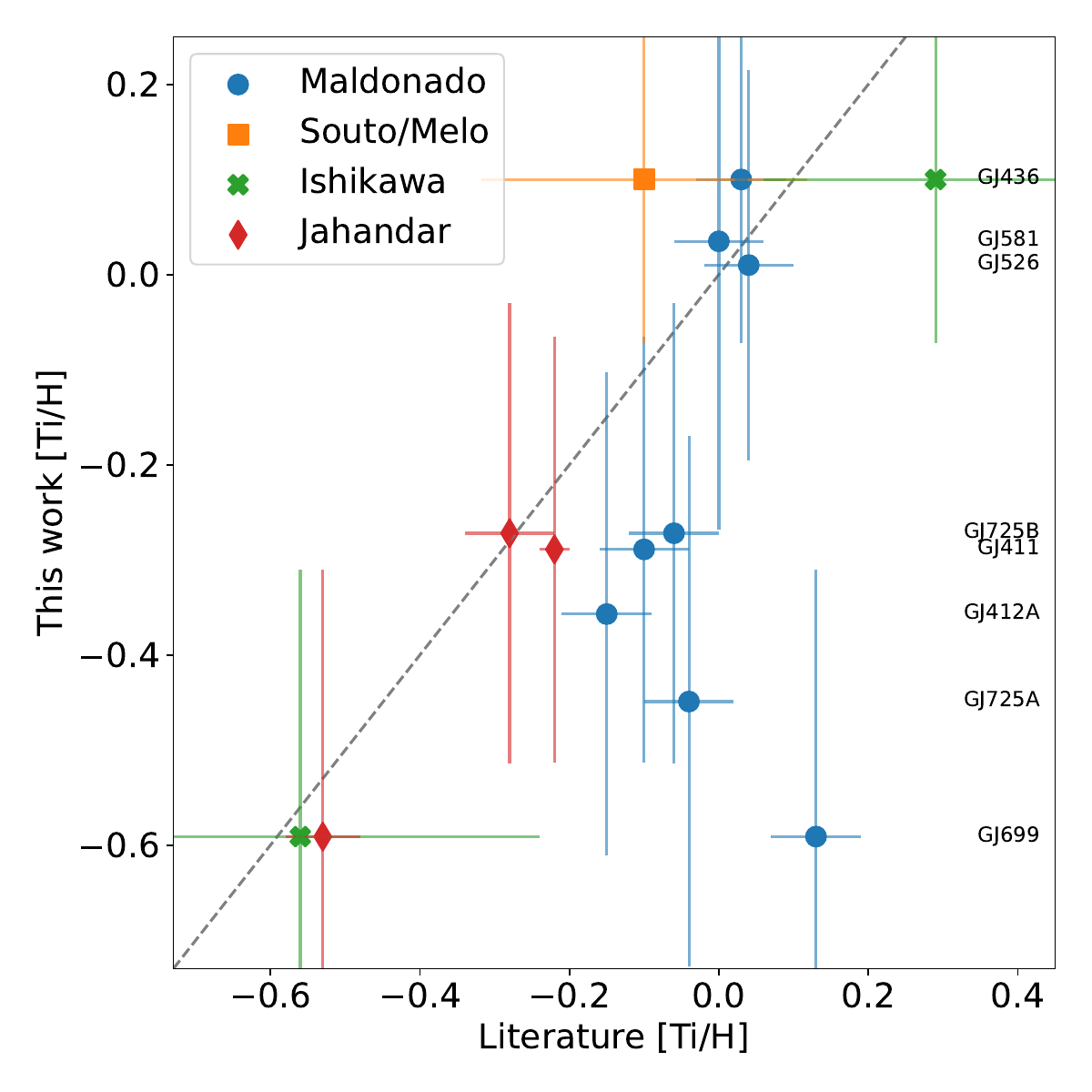}
   \includegraphics[width=5.8cm]{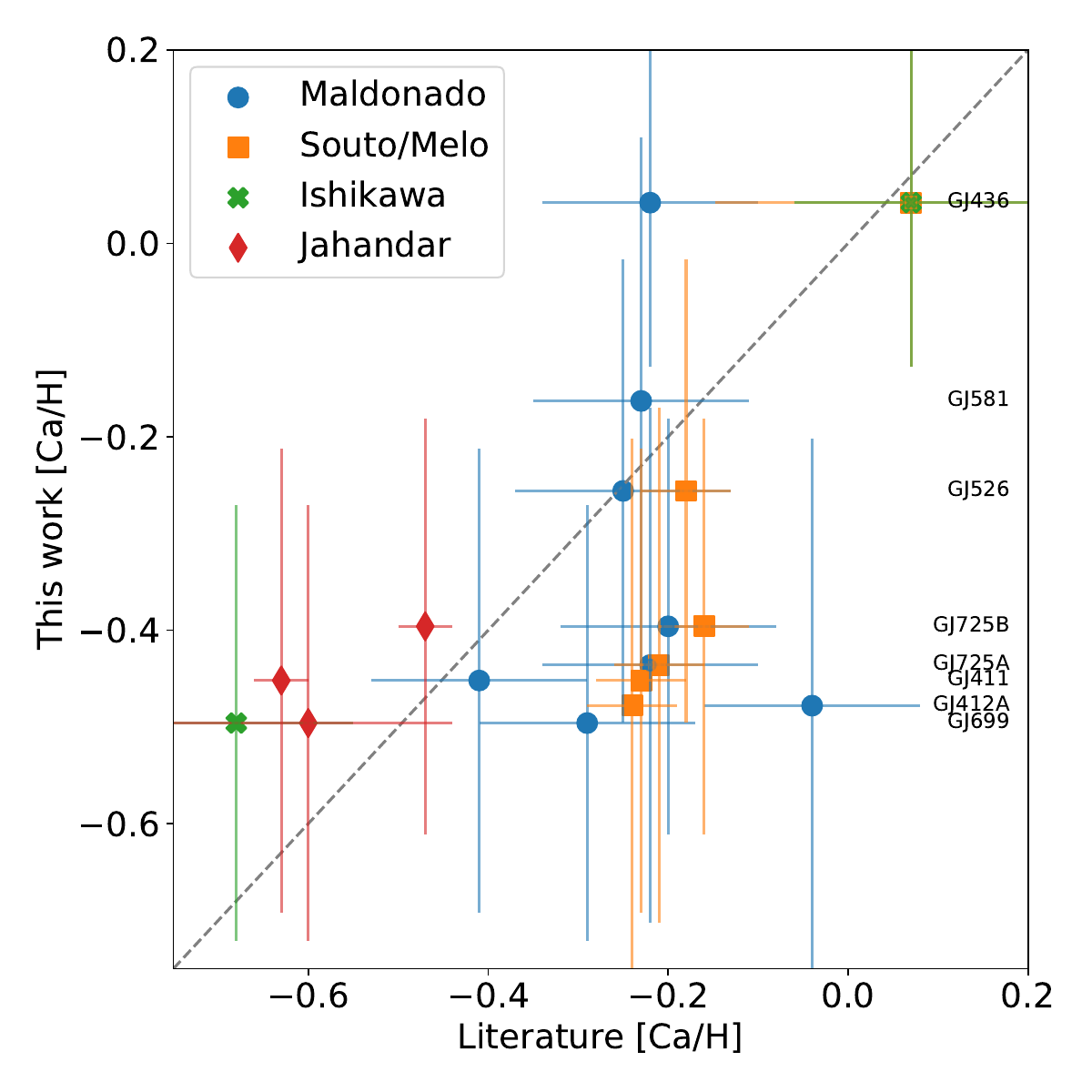}
   \caption{Abundances from this work vs abundances from the literature for stars in overlapping samples. The same star can be found on the same horizontal line with the name of the star to the right in each figure.}
   \label{fig:abund_vs_abund}
\end{figure*}

In Fig.~\ref{fig:abund_vs_abund} we compare with the overlapping sample of \citet{2022AJ_Ishikawa}, \citet{2024arXiv_Jahandar}, \citet{2020AAMaldonado}, and \citet{2022ApJSouto}. In the case of $\feh$ we are also including \citet{2015ApJMann} which we used for input metallicity.  
\citet{2022AJ_Ishikawa} used spectra from the IRD instrument mounted on the Subaru telescope. They generated synthetic spectra using MARCS atmospheric models and line lists from VALD and matched the equivalent width of the observed spectra with that of the synthetic. 
\citet{2024arXiv_Jahandar} used high resolution spectra from CFHT/SPIRou and the PHOENIX-ACES grid of synthetic spectra and determined $\teff$ and $\lgg$ prior to the abundances. \citet{2020AAMaldonado} used archival HARPS and HARPS-N data. They obtained $\teff$ and metallicity using equivalent widths (msdlines) and $\lgg$ from mass-radius relations. To get abundances they used a principal component analysis. We are grouping \citet{2022ApJSouto} and \citet{2024ApJ_Melo} together as they use the same method and spectra. They analysed APOGEE spectra using Turbospectrum and MARCS atmospheric models with the BACCHUS wrapper. They obtained $\teff$, $\lgg$, $\feh$, and oxygen and carbon abundances by using so-called $\teff$-A(O) and $\lgg$-A(O) pairs. They also obtained abundances of various atomic species by fitting individual lines. 

In Fig.~\ref{fig:abund_vs_abund} we show $\feh$ on the left, $\tih$ in the middle, and $\cah$ on the right. 
On the y-axis we show our results and on the x-axis the literature results. The error bars show the uncertainties. In each panel, the name if the star at the different horizontal levels can be found to the furthest right.

Our derived $\feh$ agree within uncertainties with \citet{2015ApJMann} (brown star symbols). One outlier is GJ~809 where we derived 0.221~dex and \citet{2015ApJMann} $-0.06$ dex. None of the other studies we are comparing with in the figure includes this star. However, \citet{Pass2019} who used CARMENES spectra and PHOENIX-ACES models obtained 0.29 dex for the same star when using the combined visual and NIR data. When comparing with the literature values for all of the other stars we mostly agree within uncertainties. The general trend is that our $\feh$ is lower than the literature values. The spread is largest when comparing with \citet{2020AAMaldonado}. Two outliers are GJ~412A and GJ~699. For GJ~412A we derived \(-0.45\)~dex and \citet{2020AAMaldonado} obtained \(-0.78\)~dex. However, \citet{2015ApJMann} obtained \(-0.37\)~dex and \citet{2022ApJSouto} \(-0.40\)~dex which we agree better with. Our derived $\feh$ for GJ~699 is significantly lower than three of the studies, \(-0.58\)~dex compared with \(-0.28\), \(-0.34\), and \(-0.40\) (\citealt{2020AAMaldonado}, \citealt{2024arXiv_Jahandar}, and \citealt{2015ApJMann}). However, \citet{2022AJ_Ishikawa} obtained \(-0.60\)~dex. \citet{Pass2019} obtained \(-0.09\) when using the combined wavelength regions, and \(-0.13\) and \(-0.14\)~dex when using the NIR or the visual alone, respectively. It is clear that this star needs further studies. We note that GJ~699 had the fewest Fe lines used in the fitting (seven).

In the middle panel we show the comparison with $\tih$. We agree within the uncertainties when comparing with \citet{2024ApJ_Melo}, \citet{2022AJ_Ishikawa}, and \citet{2024arXiv_Jahandar}. When comparing with \citet{2020AAMaldonado} we find a bigger discrepancy, especially at lower $\tih$. One outlier is the star GJ~699 where we derived $-0.591$~dex and \citet{2020AAMaldonado} $-0.130$~dex. Another example is GJ~725A. We note that these two stars show some of the largest differences in $\feh$ when comparing with \citet{2015ApJMann}. The code used in this work provides the abundance in relation to Fe and it is therefore possible that the discrepancy we see in $\tih$ is hereditary from $\feh$. In order to decipher if this is the case further study is needed, which is outside the scope if this paper.

The right panel shows the result for $\cah$ and here the spread is larger when comparing with all of the literature values. Here our uncertainties are even larger. This is most likely because we had problems fitting the cores of some of the lines as mentioned earlier. One possible explanation for this bad fit could be non-LTE effects, especially in the solar spectrum. We find a clear offset when comparing with \citet{2022ApJSouto} and \citet{2024ApJ_Melo} where our derived abundances are 0.2~dex lower than those of these authors. The largest spread is found when comparing with \citet{2020AAMaldonado}. GJ~412A is again an outlier when comparing with \citet{2020AAMaldonado} where they derived $-0.04$~dex and we derived $-0.478$~dex. \citet{2022ApJSouto} derived an abundance of $-0.24$~dex.

\begin{figure}
   \centering
   \includegraphics[width=8cm]{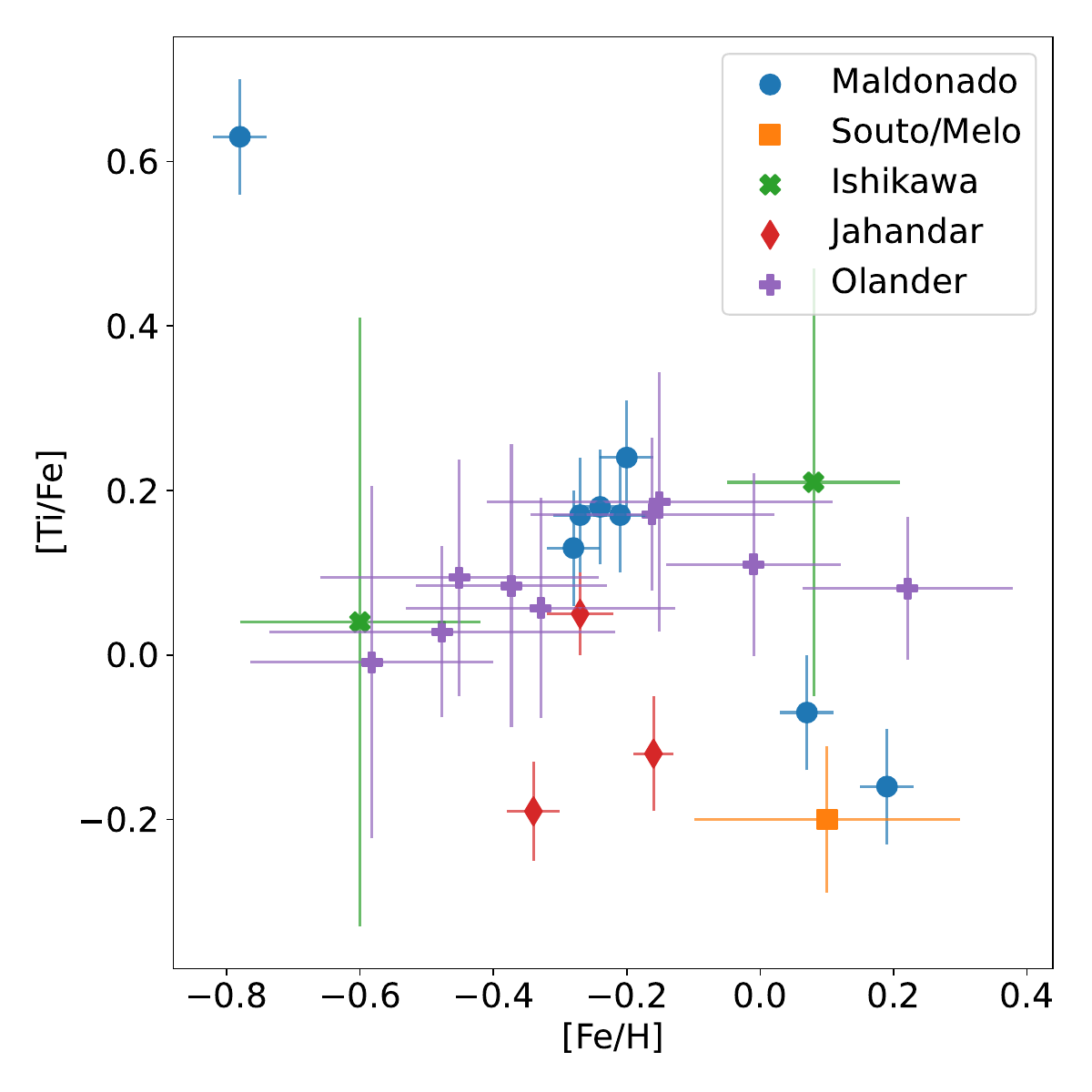}
   \includegraphics[width=8cm]{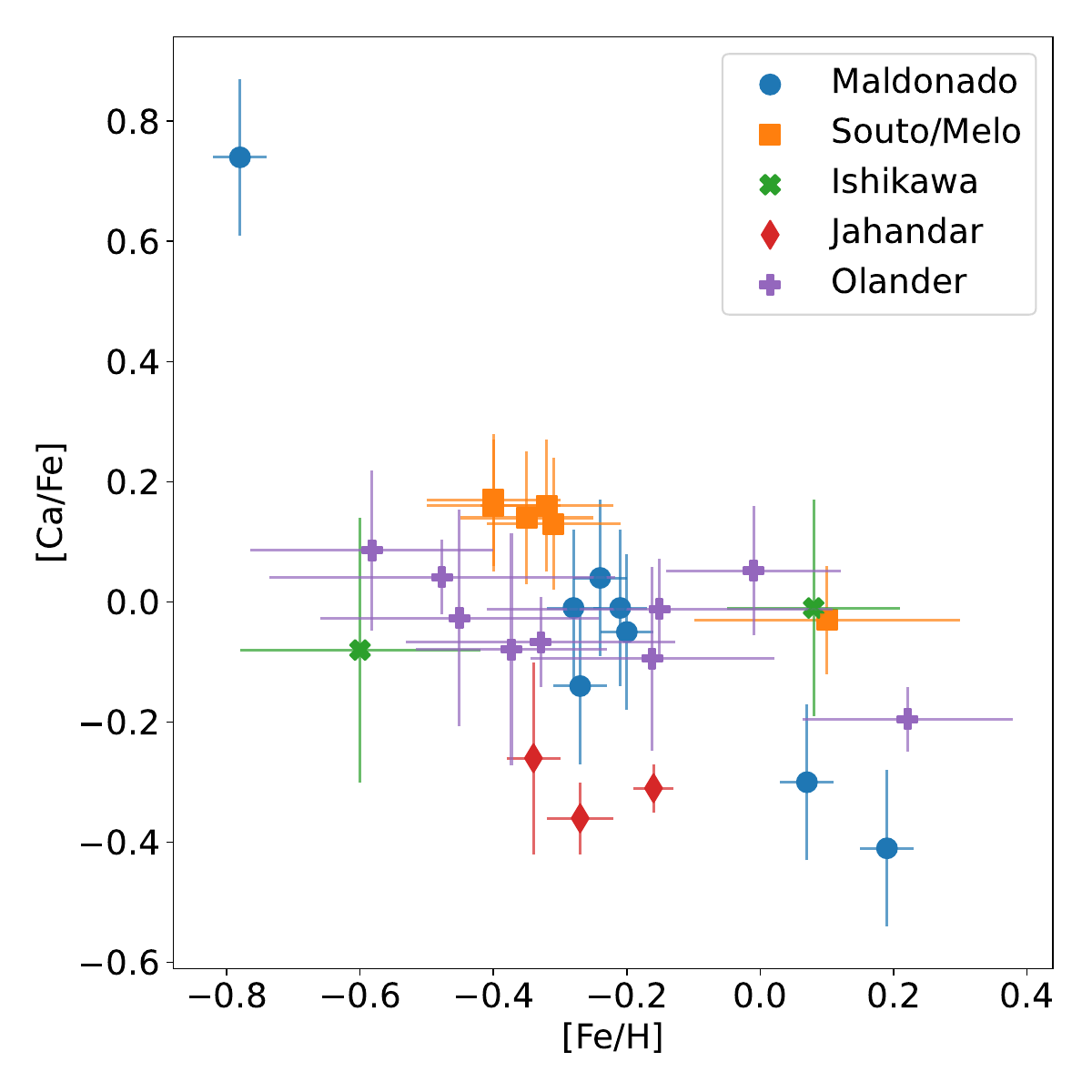}
   \caption{Abundances of Ti and Ca vs Fe for stars in the sample overlapping with studies from the literature.}
   \label{fig:abund_vs_fe}
\end{figure}

In Fig.~\ref{fig:abund_vs_fe} we plot $\tife$ and $\cafe$ against $\feh$ for all the stars in the overlapping sample with the literature studies. The symbols are the same as in Fig.~\ref{fig:abund_vs_abund}. For both $\tife$ and $\cafe$ we have a large spread and none of the studies we compare with really line up well with each other. However the sample here is too small to derive any concrete conclusion regarding trends between the compared studies.

We have one binary system in the sample, GJ~725 A and B. Binaries are expected to have the same abundances as they were formed from the same molecular cloud. For $\feh$ we differ by approximately 0.15~dex between the stars in the binary, for $\tife$ by 0.03~dex, and for $\cafe$ by 0.1~dex. Among the literature studies we are comparing with \citet{2020AAMaldonado} and \citet{2022ApJSouto} analysed both stars in the binary system. \citet{2020AAMaldonado} obtained a difference between the A and B components of 0.03~dex for $\feh$, 0.01~dex for $\tife$ and 0.05~dex for $\cafe$. \citet{2022ApJSouto} obtained a difference of 0.03~dex for $\feh$ and 0.02~dex for $\cafe$ (Ti was not measured for these stars by \citealt{2022ApJSouto}). The difference we see in abundances between the stars in the binary is an indicator of a high uncertainty in our derived abundances. For the B star our $\feh$ is close to the \citet{2022ApJSouto} value ($-0.33$~dex compared to $-0.32$~dex). It is the A star that differs ($-0.48$~dex). We note that different spectral lines were used in the fitting of the A star and B star, this can be seen in Table~\ref{tab:fitted_Fe_lines}. When we look at the $\tih$ and $\cah$ derived for the binary, the difference in abundance between the two binary stars is larger than it is for $\tife$ and $\cafe$. This indicates that it might be the iron abundance for GJ~725A that is problematic. One possible explanation for the much lower $\feh$ in the A star compared to the B star is the quality of the spectrum. As can be seen in Table~\ref{tab:sample} the SNR of GJ~725A is lower than that of GJ~725B. A more extended analysis of this binary system is needed.

\section{Discussion and conclusions}
\label{sec:discussion}
We have successfully used Turbospectrum with MARCS atmospheric models in the wrapper TSFitPy to derive abundances of Fe, Ti, and Ca for a small set of benchmark M~dwarfs. We performed a differential abundance analysis with respect to the Sun in order to minimise model dependence. The analysis was performed line-by-line and the fitted lines were carefully inspected visually. The fit of the used atomic lines are good but some of the uncertainties are high. This is partly due to problems with fringing in the observed spectra but it could also be due to the large gap between spectral type (Sun and M~dwarf).
The fringing caused an incorrect continuum placement for some of the lines and the resulting abundance for a line could therefore either be too high or too low for those particular lines. We discarded some lines because of issues with the continuum placement. 
Spectra without fringing would produce a better result with lower uncertainties.

A differential abundance analysis minimises the model dependence of the analysis. However, \citet{1989ARA&A_Gustafsson} discusses that a differential abundance analysis performs better when the analysis is done step-wise if the gap in spectral type is large. The ideal way to analyse an M~dwarf would then be to analyse the Sun (G-type), a well known K-type star relative to the Sun, and then the M~dwarf relative to the K-type star. This would also make it easier to analyse further elements, for example Si. Upon inspection of the spectra in our study it was found that the available Si lines in the wavelength region we are analysing are strong in the Sun but weak in M~dwarfs. The opposite was true for the Ti lines but in the case of Ti we found a sufficient number of suitable lines to fit. Another way to improve the differential abundance analysis would be to use spectra obtained with the same instrument for all types of stars. This would minimise instrumental effects.  

Another possible contributor to the high uncertainties is that we discarded strong lines because of a poor fit between the synthetic and the observed lines. 
The synthetic lines were either too strong or too weak and those lines were therefore discarded from the analysis because of derived abundances outside of the accepted range for that star, see Sect.~\ref{sec:Method}. The reason for the bad fit is unknown but one possible explanation could be incorrect atomic data. We used a generic line list from VALD which means that no astrophysical corrections of the $\lggf$ values were made.
Even if the effect of the line list is minimised by the differential abundance analysis the results would improve with a line list with astrophysically corrected $\lggf$ values, which was out of scope in this paper. Additionally, with a line list optimised for M~dwarfs a differential abundance analysis would not be as necessary. 

Our results show a somewhat large spread in comparison with literature values. However, there is also a spread when comparing the different literature values with each other. This can be clearly seen in Fig.~\ref{fig:abund_vs_fe}. The figure shows the overlapping sample of stars meaning that it is the same set of stars. We can see that the result from the included studies differ outside of uncertainties which indicates that the uncertainties are underestimated in many studies. The spectroscopic community needs to put more effort into analysing M~dwarfs in order to obtain fundamental parameters and abundances. We also need to explore and find the reasons for the differentiating results. Doing this would help our understanding of M~dwarfs and the physical conditions inside their atmospheres. Additionally, more abundance analysis studies for a large quantity of M~dwarfs are needed in order to draw any conclusions regarding M~dwarf abundances in connection to Galactic chemical trends. \citet{2022ApJSouto} state that their Ca abundances agree with the Galactic trend, with $\cafe$ increasing towards lower $\feh$. The $\cafe$ values derived in this work in general follow this trend. This can be seen in the lower panel of Fig.~\ref{fig:abund_vs_fe}, where our values (purple pluses) are higher at lower $\feh$ and lower at higher $\feh$. However, our sample is too small for us to say whether M~dwarfs in general follow the Galactic trend of $\cafe$.

For some of the lines included in the analysis but also for some of the discarded lines there is a small difference in line strength. An example ca be seen in the first panel in Fig.~\ref{fig:Ca_Gj436} where the observed line is stronger than the synthetic. One reason is incorrect abundances but another is departure from local thermodynamic equilibrium (non-LTE effects). This is especially true for the Sun as the conditions in the atmosphere allow for departure from LTE for a significant amount of elements. In Fig.~8 in \citet{2024ARA&A_Lind} we can see that a Ca line in the optical wavelength region is clearly affected by non-LTE effects where the 1D non-LTE line is deeper than the LTE line. We also note that the 3D LTE line is deeper than the 3D non-LTE line, which indicates that the 1D models are missing some physics. No investigation of non-LTE effects for Ca in M~dwarfs has been done to the knowledge of the authors. M~dwarfs are assumed to be in LTE but can be affected by non-LTE as shown in \citet{Olander2021}, where abundance corrections of up to 0.2~dex were found for K. Non-LTE effects for Ti were investigated in \citet{Hauschildt_1997ApJ...488..428H} who found a clear difference between LTE and non-LTE. \citet{Ti_bachelor} found an abundance correction of 0.1~dex between LTE and non-LTE. It is clear that non-LTE needs to be thoroughly investigated for M~dwarfs. 

In the future we should explore the inclusion of molecular lines when fitting in order to derive abundances of various elements. \citet{2022ApJSouto} used H$_2$O and OH molecules in order to find O abundances. It was also used to derive $\teff$ and $\lgg$. \citet{2020ApJ_Souto} found that using FeH lines does not give the same $\feh$ as using atomic Fe lines. They estimated a difference of 0.12~dex between $\feh$ from Fe~I and FeH lines in the H band. In our analysis of the star GJ~699 we found it difficult to derive $\feh$ as the Fe~I lines were weak. When inspecting Fig.~2 in \citet{2020ApJ_Souto} we can see that the stars with lower $\teff$ and $\feh$ have a better agreement between $\feh$ obtained from Fe~I lines and from FeH lines. This means that if we have a first estimate of the iron abundance and the star is cool we could use FeH lines as well as Fe~I lines. This could possibly lower the uncertainty of $\feh$ in cool, low-metallicity stars such as GJ~699. A more extended investigation is needed in order to test this. FeH lines could also be used to obtain other parameters such as $\teff$, as shown in \citet{Lind2016,Lind2017}.


\begin{acknowledgements}
We thank the anonymous referee for valuable comments that helped to improve the manuscript.
T.O., U.H., and O.K. acknowledge support from the Swedish National Space Agency (SNSA/Rymdstyrelsen).
O.K. also acknowledges funding by the Swedish Research Council (grant agreements no. 2019-03548 and 2023-03667).
Based on observations made with the Italian Telescopio Nazionale Galileo (TNG) operated on the island of La Palma by the Fundación Galileo Galilei of the INAF (Istituto Nazionale di Astrofisica) at the Spanish Observatorio del Roque de los Muchachos of the Instituto de Astrofisica de Canarias.
This work has made use of the VALD database, operated at Uppsala University and the University of Montpellier.
We thank Tanya Ryabchikova for her invaluable work with compiling and assessing atomic data for the VALD database, and Bertrand Plez for providing molecular line lists.
\end{acknowledgements}


\bibliographystyle{aa}
\bibliography{References}


\begin{appendix}

\begin{table*}[h!]
\section{Line data and lines fitted for each star}
\label{append:fitted_lines}
In this appendix we present the atomic data for the spectral lines used in the analysis in Table~\ref{tab:linelist}. We also specify which lines were fitted for each star. Table~\ref{tab:fitted_Fe_lines} shows the Fe lines, Table~\ref{tab:fitted_Ti_lines} shows the Ti lines, and Table~\ref{tab:fitted_Ca_lines} shows the Ca lines. In the latter three tables, the leftmost column indicates the rest wavelength in air of each line in \AA. There is one column for each star in the analysis where an ``X'' marks if a line was used or not.
These tables are referred to in Sect.~\ref{sec:method:linedata}.

\medskip

\caption{Atomic data for the spectral lines used in the analysis.}
\label{tab:linelist}
\centering
\begin{tabular}{lrrrrrl}
\hline
\noalign{\smallskip}
Species & $\lambda_{air}$ & $E_{\rm low}$ & $\log gf$ & $\log\gamma_{\rm rad}$ & $\log\gamma_{\rm Waals}$ & References \\
 & [\AA] & [eV] & & & & \\
\noalign{\smallskip}
\hline
\noalign{\smallskip}
Fe & 10340.885 &  2.1979 &  -3.577 & 7.250 & -7.800 & K14 \\
Ca & 10343.819 &  2.9325 &  -0.300 & 8.500 & -7.480 & K07 \\
Fe & 10395.794 &  2.1759 &  -3.393 & 7.260 & -7.800 & K14 \\
Ti & 10396.800 &  0.8484 &  -1.539 & 5.140 & -7.810 & K16 \\
Fe & 10423.027 &  2.6924 &  -3.616 & 6.130 & -7.810 & K14 \\
Fe & 10423.743 &  3.0713 &  -2.918 & 6.740 & -7.790 & K14 \\
Fe & 10435.355 &  4.7331 &  -1.945 & 8.350 & -7.550 & K14 \\
Fe & 10469.652 &  3.8835 &  -1.184 & 6.830 & -7.820 & K14 \\
Ti & 10496.116 &  0.8360 &  -1.651 & 5.130 & -7.810 & K16 \\
Fe & 10532.234 &  3.9286 &  -1.480 & 6.820 & -7.820 & K14 \\
Ca & 10558.425 &  4.5541 &  -0.731 & 7.330 & -6.960 & K07 \\
Fe & 10577.139 &  3.3014 &  -3.136 & 7.960 & -7.770 & K14 \\
Ti & 10584.634 &  0.8259 &  -1.775 & 5.130 & -7.810 & K16 \\
Ti & 10607.716 &  0.8484 &  -2.697 & 5.130 & -7.810 & K16 \\
Fe & 10616.721 &  3.2671 &  -3.127 & 7.990 & -7.780 & K14 \\
Ti & 10661.623 &  0.8181 &  -1.915 & 5.130 & -7.810 & K16 \\
Ti & 10732.865 &  0.8259 &  -2.515 & 5.130 & -7.810 & K16 \\
Ti & 10774.866 &  0.8181 &  -2.666 & 5.130 & -7.810 & K16 \\
Fe & 10783.050 &  3.1110 &  -2.567 & 6.740 & -7.790 & K14 \\
Fe & 10818.274 &  3.9597 &  -1.948 & 6.820 & -7.820 & K14 \\
Ca & 10827.013 &  4.8770 &  -0.300 & 8.220 & -6.930 & K07 \\
Ca & 10838.970 &  4.8775 &   0.238 & 8.400 & -7.590 & K07 \\
Ca & 10861.582 &  4.8767 &  -0.343 & 8.390 & -7.590 & K07 \\
Fe & 10863.518 &  4.7331 &  -0.895 & 8.350 & -7.550 & K14 \\
Fe & 10884.262 &  3.9286 &  -1.925 & 6.830 & -7.820 & K14 \\
Fe & 10896.299 &  3.0713 &  -2.694 & 6.880 & -7.790 & K14 \\
Fe & 11607.572 &  2.1979 &  -2.009 & 7.160 & -7.820 & BWL, K14 \\
Ca & 11759.570 &  4.5313 &  -0.878 & 7.500 & -7.090 & K07 \\
Ca & 11769.345 &  4.5322 &  -1.011 & 7.500 & -7.090 & K07 \\
Fe & 11884.083 &  2.2227 &  -2.083 & 7.160 & -7.820 & BWL, K14 \\
Ca & 11955.955 &  4.1308 &  -0.849 & 8.010 & -7.300 & K07 \\
Fe & 12227.112 &  4.6070 &  -1.368 & 8.430 & -7.540 & K14 \\
Fe & 12342.916 &  4.6382 &  -1.463 & 8.420 & -7.540 & K14 \\
Ti & 12600.277 &  1.4432 &  -2.320 & 6.810 & -7.790 & LGWSC, Sal12a, K16 \\
Fe & 12807.152 &  3.6398 &  -2.452 & 8.080 & -7.750 & K14 \\
Ti & 12811.478 &  2.1603 &  -1.390 & 7.990 & -7.750 & LGWSC, Sal12a, K16 \\
Ca & 12816.045 &  3.9104 &  -0.765 & 8.280 & -7.520 & K07 \\
Ti & 12821.672 &  1.4601 &  -1.190 & 6.810 & -7.790 & LGWSC, Sal12a, K16 \\
Ca & 12823.867 &  3.9104 &  -0.997 & 8.280 & -7.520 & K07 \\
Fe & 12824.859 &  3.0176 &  -3.835 & 6.070 & -7.810 & K14 \\
Ca & 12827.059 &  3.9104 &  -1.478 & 8.280 & -7.520 & K07 \\
Ti & 12831.445 &  1.4298 &  -1.490 & 6.820 & -7.790 & LGWSC, Sal12a, K16 \\
Fe & 12840.574 &  4.9556 &  -1.329 & 8.710 & -7.550 & K14 \\
Fe & 12879.766 &  2.2786 &  -3.458 & 7.170 & -7.820 & BWL, K14 \\
Ca & 12909.070 &  4.4300 &  -0.224 & 7.770 & -7.710 & K07 \\
Ti & 13011.897 &  1.4432 &  -2.270 & 6.820 & -7.790 & LGWSC, Sal12a, K16 \\
Ca & 13033.554 &  4.4410 &  -0.064 & 7.770 & -7.710 & K07 \\
Ca & 13057.885 &  4.4410 &  -1.092 & 7.770 & -7.710 & K07 \\
\noalign{\smallskip}
\hline
\end{tabular}
\tablefoot{All lines are from neutral atomic species (i.e., Fe~I, Ca~I, Ti~I). $\lambda$ \ldots wavelength, $E_{\rm low}$ \ldots lower level energy, $\log gf$ \ldots logarithm (base 10) of the product of the oscillator strength of the transition and the statistical weight of the lower level, $\log\gamma_{\rm rad}$ \ldots logarithm of the radiative damping width in units of rad~s$^{-1}$, $\log\gamma_{\rm Waals}$ \ldots logarithm of the van der Waals broadening width per unit perturber number density at 10\,000~K in units of rad s$^{-1}$ cm$^{3}$. Unknown damping parameters are set to zero.}
\tablebib{
BWL    \ldots \citet{BWL},
K07    \ldots \citet{K07},
K14    \ldots \citet{K14},
K16    \ldots \citet{K16},
LGWSC  \ldots \citet{LGWSC},
Sal12a \ldots \citet{Sal12a}.
}
\end{table*}

\begin{table*}[h!]
\caption{Fitted Fe~I lines.}
\label{tab:fitted_Fe_lines}
\centering
\begin{tabular}{lccccccccc}
\noalign{\smallskip}
\hline
\noalign{\smallskip}
$\lambda_{air}$~[\AA] & GJ~411 & GJ~412A & GJ~436 & GJ 526 & GJ 581 & GJ 699 & GJ 725A & GJ 725B & GJ 809 \\
\noalign{\smallskip}
\hline
\noalign{\smallskip}
10340.885 & X & X & X & X & X &  &  &  & X \\
10395.794 & X & X & X & X & X &  & X &  & X \\
10423.027 & X &  & X & X & X & X & X & X & X \\
10423.743 &  &  & X &  & X &  &  & X & X \\
10435.355 & X & X & X & X & X & X &  &  & X \\
10469.652 & X & X & X & X &  & X & X & X & X \\
10532.234 & X & X & X & X & X &  &  & X & X \\
10577.139 &  &  &  & X & X &  &  & X & X \\
10616.721 &  &  &  & X &  &  &  &  & X \\
10783.050 & X & X & X & X & X & X & X & X & X \\
10818.274 &  &  &  & X &  &  &  &  & X \\
10863.518 &  &  &  &  &  &  & X &  &  \\
10884.262 &  &  &  & X &  &  &  &  & X \\
10896.299 & X & X & X & X & X &  & X &  & X \\
11607.572 &  &  &  &  & X &  &  &  &  \\
11884.083 & X & X & X &  & X & X & X & X & X \\
12227.112 & X &  &  & X &  &  & X & X & X \\
12342.916 &  &  &  & X &  &  &  &  & X \\
12807.152 & X & X & X & X & X & X & X & X & X \\
12824.859 &  &  &  & X &  &  &  &  & X \\
12840.574 &  &  &  &  &  &  &  &  & X \\
12879.766 & X & X & X & X & X & X & X & X &  \\
\noalign{\smallskip}
\hline           
\end{tabular}
\end{table*}

\begin{table*}[h!]
\caption{Fitted Ti~I lines.}
\label{tab:fitted_Ti_lines}
\centering
\begin{tabular}{lccccccccc}
\noalign{\smallskip}
\hline
\noalign{\smallskip}
$\lambda_{air}$~[\AA] & GJ 411 & GJ 412A & GJ 436 & GJ 526 & GJ 581 & GJ 699 & GJ 725A & GJ 725B & GJ 809 \\
\noalign{\smallskip}
\hline
\noalign{\smallskip}
10396.800 & X & X & X & X & X & X & X & X & X \\
10496.116 & X & X & X & X & X & X & X & X & X \\
10584.634 &  &  &  &  & X &  &  &  &  \\
10607.716 & X & X & X & X & X & X & X & X & X \\
10661.623 & X & X & X & X & X & X &  & X & X \\
10732.865 & X & X & X & X & X & X & X & X & X \\
10774.866 & X & X & X & X & X & X &  & X & X \\
12600.277 & X & X & X & X & X & X & X & X & X \\
12811.478 & X & X & X & X & X & X & X & X & X \\
12821.672 & X & X & X & X & X & X & X & X & X \\
12831.445 & X & X & X & X & X & X & X & X & X \\
13011.897 &  &  & X &  &  & X & X &  &  \\
\noalign{\smallskip}
\hline           
\end{tabular}
\end{table*}

\begin{table*}[h!]
\caption{Fitted Ca~I lines.}
\label{tab:fitted_Ca_lines}
\centering
\begin{tabular}{lccccccccc}
\noalign{\smallskip}
\hline
\noalign{\smallskip}
$\lambda_{air}$~[\AA] & GJ 411 & GJ 412A & GJ 436 & GJ 526 & GJ 581 & GJ 699 & GJ 725A & GJ 725B & GJ 809 \\
\noalign{\smallskip}
\hline
\noalign{\smallskip}
10343.819 & X & X & X & X & X & X & X & X & X \\
10558.425 &  &  & X &  & X &  &  &  &  \\
10827.013 &  & X &  & X &  &  &  &  &  \\
10838.970 & X & X & X & X & X &  & X &  & X \\
10861.582 &  &  &  & X & X &  &  & X & X \\
11759.570 & X &  &  &  &  &  &  &  &  \\
11769.345 &  &  &  &  & X &  & X & X &  \\
11955.955 & X & X & X & X & X &  & X & X & X \\
12816.045 &  &  &  &  &  & X &  &  &  \\
12823.867 & X & X & X & X & X & X & X & X & X \\
12827.059 & X &  & X & X & X &  & X & X & X \\
12909.070 & X & X & X & X & X & X & X & X & X \\
13033.554 & X & X & X & X & X & X & X & X & X \\
13057.885 &  &  &  & X &  & X &  & X & X \\
\noalign{\smallskip}
\hline           
\end{tabular}
\end{table*}

\end{appendix}

\end{document}